\documentclass[aps,prl,twocolumn,groupedaddress]{revtex4}

\usepackage{amssymb}
\usepackage{amsmath}
\usepackage{bm}
\usepackage{graphicx}
\usepackage{dcolumn}
\usepackage{longtable}
\usepackage{color}

\begin{document}


\title{Anisotropic magnetic excitations from single-chirality antiferromagnetic state in Ca-kapellasite}

\author{Y.~Ihara}
\email{yihara@phys.sci.hokudai.ac.jp}
\affiliation{Department of Physics, Faculty of Science, Hokkaido University, Sapporo 060-0810, Japan}
\author{K.~Arashima}
\affiliation{Department of Condensed Mattter Physics, Graduate School of Science, Hokkaido University, Sapporo 060-0810, Japan}
\author{H. Yoshida}
\affiliation{Department of Physics, Faculty of Science, Hokkaido University, Sapporo 060-0810, Japan}
\author{M. Hirata}
\affiliation{Institute for Materials Research, Tohoku University, Sendai 980-8577, Japan}
\author{T. Sasaki}
\affiliation{Institute for Materials Research, Tohoku University, Sendai 980-8577, Japan}
\date{\today}

\begin{abstract}
We present a $^{35}$Cl NMR study for spin $S=1/2$ perfect kagome antiferromagnet Ca-kapellasite (CaCu$_{3}$(OH)$_{6}$Cl$_{2}\cdot$0.6H$_{2}$O) with a magnetic transition at $T^{\ast}=7.2$ K. 
The static magnetic structure in the ground state has been determined to be a chirality-ordered $Q=0$ state, 
which is selected by a finite Dzyaloshinskii-Moriya interaction. 
The low-energy magnetic excitations in the ordered state are investigated by the nuclear spin-lattice relaxation rate measurement. 
We detect a weakly temperature dependent contribution in the magnetic fluctuations perpendicular to the kagome plane 
in addition to the dispersive spin-wave contribution in the kagome plane. 
The low-energy magnetic excitations from the coplanar spin structure are attributed to the zero mode originating from the flat band in this kagome antiferromagnet. 
\end{abstract} 

\maketitle

The ground state of quantum magnets with spin $S =1/2$ has been intensively discussed 
because the quantum fluctuations destabilize the canonical magnetic ordering and give rise to a new state of matter known as the spin liquid state. \cite{balents-Nature464}
This prominent quantum effect emerges 
when the magnetic ordering is suppressed at very low temperatures by the competing interactions among several spins. 
One commonly recognized example is the geometrical frustration effect for antiferromagnetically interacting spins on a triangular lattice. 
Magnets with triangle-based spin configuration such as kagome and pyrochlore networks for two-dimensional (2-D) and three-dimensional (3-D) structures 
are also candidates for experimental realization of the spin liquid state. 
In fact, the spin liquid state has been observed in herbertsmithite (2-D) \cite{shores-JACS127, norman-RMP88, han-Nature492, imai-PRL100, fu-Science350, jeong-PRL107} 
and Yb$_2$Ti$_2$O$_7$ (3-D). \cite{ross-PRX1, chang-natcom3, tokiwa-natcom7}
A recently found new candidate for a spin liquid material is the Kitaev ferromagnet, 
for which theory suggests that the spin-orbit coupled anisotropic frustration leads to a spin liquid ground state. \cite{kitaev-anphy321, nasu-NatPhy12}
In the spin liquid state the spin-spin interactions extend to the distance, yet the spins maintain the dynamics. 
As a result the elemental excitation can be fractionalized and 
fermionic spinon excitations were observed in $\alpha$-RuCl$_3$ \cite{sandilands-PRL114, kasahara-Nature559} and Ir based magnets \cite{singh-PRL108,kitagawa-Nature554}. 
In the case of Yb$_2$Ti$_2$O$_7$ the monopole excitations were observed. \cite{tokiwa-natcom7}
Experimental observation of fractionalized magnetic excitations is not only striking evidence for the spin liquid state, but also 
a characteristic potentially applicable to the quantum computation. 
Although fractional magnetic excitations were predicted only for the spin liquid state without any magnetic ordering, 
in kagome antiferromagnets low-energy magnetic excitations can be generated even in the magnetically ordered state 
because of the macroscopic degeneracy left in the ground state. 

Magnetic excitations in the ordered state are generally understood as the bosonic magnons
which can be perfectly explained by spin-wave theory. \cite{yildirim-PRB73, matan-PRL96}
In the kagome antiferromagnet, however, a flat band known as the {\it zero mode} appears at zero energy in the magnon dispersion for a coplanar magnetic state \cite{chalker-PRL68, harris-PRB45} 
and its effect on the magnetic properties have not been revealed because the flat band can infinitely populate low-energy excitations. \cite{chernyshev-PRB92}
As the zero mode is the spin excitations perpendicular to the kagome plane, 
we could interpret that the spin component perpendicular to the kagome plane is disordered, while the in-plane components are ordered. 
Experimental input is crucially important to advance understanding about low-energy excitations in $S=1/2$ kagome antiferromagnets, but so far no experimental results have been reported 
due to the lack of suitable materials that show a regular magnetic ordering on the perfect kagome network. 

\begin{figure}[tbp]
\begin{center}
\includegraphics[width=7.5cm]{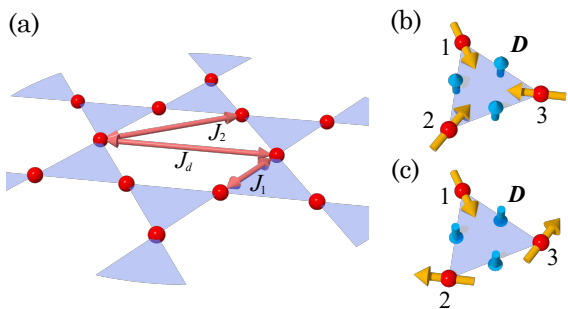}
\end{center}
\caption{
(a) Magnetic kagome network of Ca-kapellasite. 
In the kepellasite family $J_d$ across the hexagon is finite in addition to the nearest and next nearest interactions $J_1$ and $J_2$. 
The short blue arrows are the $\bm{D}$ vectors for Dzyaloshinskii-Moriya interaction. 
Spin configuration for upward and downward $\bm{D}$ vectors are shown in (b) and (c). 
Depending on the $\bm{D}$ vector directions, positive (b) and negative (c) chirality spin configurations will be selected.
} 
\label{Fig1}
\end{figure}

In this study, we focus on a 2-D kagome antiferromagnet Ca-kapellasite (CaCu$_{3}$(OH)$_{6}$Cl$_{2}\cdot$0.6H$_{2}$O), \cite{yoshida-JPSJ86}
which has a magnetically perfect kagome network of $S=1/2$ Cu spins. 
The magnetism in Ca-kapellasite is understood on the basis of the $J_1-J_2-J_d$ model, \cite{fak-PRL109, kermarrec-PRB90, boldrin-PRB91, bieri-PRB92, iqbal-PRB92}
in which the long range interaction across the hexagon $J_d$ is finite in addition to the nearest and next nearest interactions $J_1$ and $J_2$. [Fig.~\ref{Fig1} (a)]
Within the family of kapellasite materials \cite{colman-CM20,colman-CM22} , Ca-kapellasite is the most interesting because of the antiferromagnetic $J_1$. 
The ground states for antiferromagnetic $J_1$ were theoretically investigated only from the classical limit without quantum fluctuation effects. \cite{messio-PRB83}
In the phase diagram $\sqrt{3}\times \sqrt{3}$ and $Q=0$ ground states reside very close to each other
when $J_d$ is small, which is the case for Ca-kapellasite.
The ground state in the real material will be selected by further interactions, such as Dzyaloshinskii-Moriya (DM) interaction $\bm{D}\cdot \bm{S}_i \times \bm{S}_j$
which is intrinsic for the kagome network. \cite{elhajal-PRB66}
The DM vector $\bm{D}$ perpendicular to the kagome plane stabilizes the coplanar magnetic state [Figs.~\ref{Fig1} (b), (c)] opening a gap for the zero mode. \cite{chernyshev-PRB92}

Previous bulk property measurements \cite{yoshida-JPSJ86} revealed a weak magnetic ordering at $T^{\ast}=7.2$ K from the peaks in the in-plane magnetization and the heat capacity. 
From the $^{35}$Cl NMR experiment for a powdered sample a peak in the spin-lattice relaxation rate $1/T_{1}$ at $T^{\ast}$ and spectrum broadening below $T^{\ast}$ were observed 
as the evidence for static magnetic ordering. \cite{ihara-PRB96}
In the ordered state, the dispersive magnon excitations were observed as the $T^2$ term in the heat capacity and the $T^3$ term in $1/T_{1}$. 
In addition to these magnon contributions Ca-kapellasite has nontrivial magnetic excitations which introduce $T$-linear temperature dependence both in the heat capacity and $1/T_{1}$. 
As this temperature dependence is reminiscent of a Fermi liquid state in conductors, 
one would invoke a fermionic nature for the low-energy excitations. 

To identify the origin of nontrivial magnetic excitations, we performed $^{35}$Cl NMR experiments on crystalline samples. 
From the NMR spectra and $1/T_1$ measurements in field applied parallel ($H_{||}$) and perpendicular ($H_{\perp}$) to the kagome plane, 
we found that the ordered state is highly anisotropic due to the coplanar magnetic structure. 
Moreover, we revealed that the magnetic excitations are also anisotropic and that the weakly temperature dependent contribution was observed only in the $c$ axis component. 
We discuss the zero mode in kagome antiferromagnets as a possible origin of the orientation ordered magnetic excitations. 

The $^{35}$Cl NMR experiment was performed for a single crystal with a typical dimension of $3\times 2 \times 0.5$ mm$^3$. 
The sample was mounted on a single axis rotator, with which the field direction can be tuned from the $a$ axis ($H_{||})$ to the $c$ axis $(H_{\perp}$). 
The NMR spectra were obtained by a single Fourier transformation at fixed field when the linewidth is narrow at high temperatures, 
and by recombining several Fourier transformations during the field sweep for the broad spectra at low temperatures. 
The peak frequency/field and their standard error were estimated by fitting the obtained spectrum with a Gaussian function. 
For the high field NMR experiment, we used the high $T_c$ superconducting magnet in the Institute for Materials Research, Tohoku University.

\begin{figure}[tbp]
\begin{center}
\includegraphics[width=7.5cm]{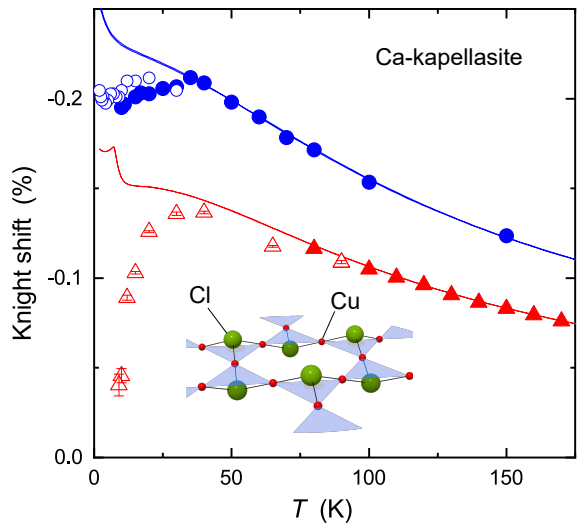}
\end{center}
\caption{
$^{35}$Cl NMR shift measured in field perpendicular (blue balls) and parallel (red triangles) to the kagome plane. 
Open (filled) symbols are Knight shift determined by the field (frequency) sweep NMR spectra. 
Consistent results were obtained for both methods at the overlapping temperatures. 
The solid lines are the temperature dependence of uniform susceptibility \cite{yoshida-JPSJ86} multiplied by the hyperfine coupling constants. 
Note that the vertical axis has negative sign because of the negative hyperfine coupling constants. 
Inset shows the Cl sites. 
Crystallographically equivalent Cl sites are located above and below the Cu triangles. 
}
\label{KnightShift}
\end{figure}

Figure \ref{KnightShift} shows the temperature dependence of the Knight shift $K$. 
Temperature-independent term was subtracted from the total shift. 
The solid lines in Fig.~\ref{KnightShift} are the uniform susceptibility $\chi$ multiplied by the hyperfine coupling constants for $H_{||} $ and $H_{\perp}$, 
which are $A_{||}=-267(3)$ mT/$\mu_B$ and $A_{\perp}=-370(3)$ mT/$\mu_B$ \cite{sup}.
At high temperatures above $30$ K $\chi$ follows $K$ as expected for a paramagnetic state. 
At low temperatures $K$ deviates from $\chi$ due to the formation of short-range ordering structure. 
The in-plane Knight shift $K_{||}$ goes down to zero around $T^{\ast}$, 
suggesting a $120$ degrees structure in the ordered state, for which the transferred hyperfine fields are canceled out at the Cl sites on the trigonal axis. (inset of Fig.~\ref{KnightShift}) 
Remarkably, the out-of-plane Knight shift $K_{\perp}$ levels off around $T^{\ast}$ and remains finite even in the ordered state. 
We can exclude any extrinsic magnetization from impurities and defects as the source of finite $K_{\perp}$
because such contributions would not shift the peak positions but increase the linewidth of the NMR spectrum. 
The finite $K_{\perp}$ evidences that the $c$-axis susceptibility is finite and thus the low-energy magnetic excitations survive partially in the antiferromagnetically ordered state. 
We will discuss the dynamics of this low-energy mode later. 

\begin{figure}[tbp]
\begin{center}
\includegraphics[width=7.5cm]{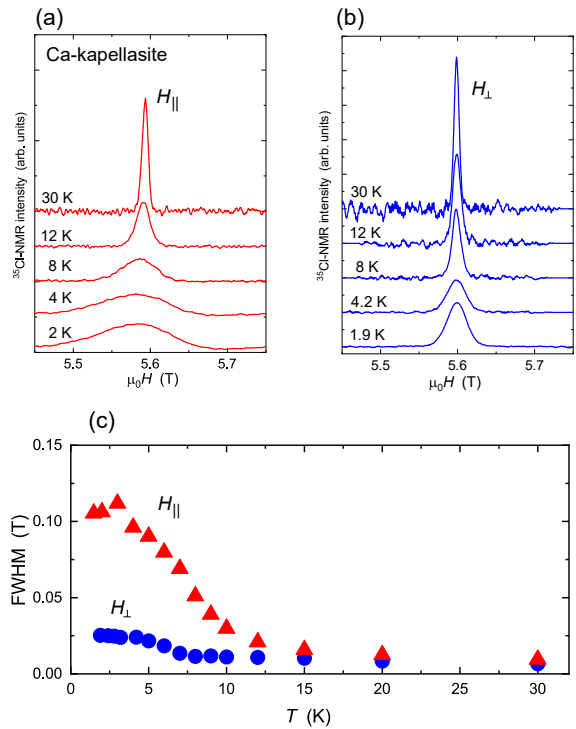}
\end{center}
\caption{
Temperature dependence of $^{35}$Cl NMR spectra obtained in (a) $H_{||}$ and (b) $H_{\perp}$. 
Spectra at each temperature are vertically shifted for visibility. 
Clear spectrum broadening was observed only in $H_{||}$, which leads us to suggest a coplanar magnetic structure. 
(c) Temperature dependence of the full width at half maximum (FWHM). 
The onset of spectrum broadening is slightly higher than $T^{\ast}$ for $H_{||}$. 
}
\label{NMRspectra}
\end{figure}

Anisotropic behavior was also observed in the linewidth. 
As shown in Fig.~\ref{NMRspectra} (a), (b), the emergent line broadening associated with magnetic transition was observed only in $H_{||}$. 
The small increase in the linewidth for $H_{\perp}$ further confirms that the finite $c$-axis susceptibility does not originate from magnetic impurities. 
In the temperature dependence of full width at the half maximum [FWHM, Fig.~\ref{NMRspectra} (c)] the onset of the line broadening was observed slightly above $T^{\ast}$ in $H_{||}$. 
The precursor to the magnetic transition is ascribed to the growth in the coherent length for interactiong spins. 
In the ordered state a broad spectrum without apparent structure was observed. 
In contrast, in $H_{\perp}$ magnetic line broadening at the lowest temperature is approximately $12$ mT, which is $1/8$ of that in $H_{||}$. 
The finite line broadening in $H_{\perp}$ can be ascribed to a field misalignment of less than $7$ degrees. 
From these results we can conclude that the internal fields at the Cl site orient parallel to the kagome plane, and its amplitude is distributed. 

To explain the reduction of $K_{||}$ from 30 K to $T^{\ast}$ we assume a coplanar $120$ degrees spin structure, where all spins are in the kagome plane. 
Then, we can consider two different spin configurations; the positive- and negative-chirality spin configurations, depending on the direction of the spin rotation 
as shown in Fig.~\ref{Fig1} (b), (c). 
In the $Q=0$ magnetic structure on the kagome network all the Cu triangles possess the same chirality, 
whereas in the $\sqrt{3}\times\sqrt{3}$ state both chirality are arranged alternatively. 
As the transferred hyperfine fields are canceled for both spin configurations, 
the spectrum broadening in $H_{||}$ is caused purely by dipole fields. 
The dipole fields at the Cl site have only the $c$-axis component for the positive chirality, and in-plane component for the negative chirality. 
The details of the dipole field calculation are given in the supplemental material. \cite{sup} 
As our experimental result unveils that the internal fields appear only along the kagome plane, 
we conclude that all the spins form the negative-chirality spin configuration. 
Thus, we suggest that the negative-chirality $Q=0$ magnetic structure is realized in Ca-kapellasite. 

The chirality of spins on a triangle is selected by the DM interaction, 
namely the $\bm{D}$ vector perpendicular to the triangular plane, \cite{elhajal-PRB66}
whereas the in-plane $\bm{D}$ component introduces a non-coplanar spin structure. 
The coplanar and negative-chirality magnetic structure suggests that $\bm{D}$ is perpendicular to the kagome plane and pointing down as Fig.~\ref{Fig1} (c).
The thermal Hall effect was also consistently explained by the negative-chirality $Q=0$ state. \cite{doki-PRL121}
The negative-chirality $Q=0$ state was also suggested for a sister compound Cd-kapellasite \cite{okuma-PRB95}, and  YCu$_{3}$(OH)$_{6}$Cl$_{3}$. \cite{zorko-PRB100}
In Cd-kapellasite the abrupt increase in susceptibility above the magnetic ordering temperature was explained as the precursor to the $Q=0$ 
magnetic state from symmetry considerations. 
A similar upturn was also observed in Ca-kapellasite. \cite{yoshida-JPSJ86}

It should be noted that the $Q=0$ state has a freedom of global spin rotation. 
If the spin directions were locked to a certain direction with respect to the external field orientation, 
only two kinds of antiparallel dipole fields would be generated at the Cl site, which results in a two-peak NMR spectrum. 
From our results, however, a broad NMR spectrum without any structure was observed as shown in Fig.~\ref{NMRspectra} (a). 
We thus suggest that $Q=0$ state forms a domain to induce distributed internal fields at the Cl sites.

\begin{figure}[tbp]
\begin{center}
\includegraphics[width=8.5cm]{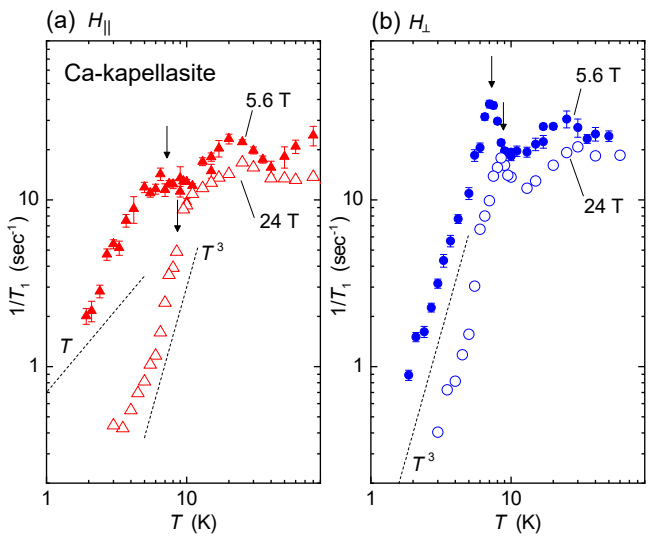}
\end{center}
\caption{
Temperature dependence of $1/T_{1}$ in (a) $H_{||}$ and (b) $H_{\perp}$. 
For both figures filled (open) symbols are the result obtained in $5.6$ T ($24$ T). 
$T^{\ast}$ at each field is determined by the peaks in $H_{\perp}$ and pointed by arrows.
The dashed lines indicate the slope for $T$-linear and $T^3$ temperature dependence. 
In $H_{\perp}$ $1/T_{1}$ shows a peak at $T^{\ast}$ and $T^3$ behavior was observed below $T^{\ast}$. 
The temperature dependence is weaker in $H_{||}$, namely at low fields and low temperatures. 
}
\label{relaxation}
\end{figure}

Next we measured the nuclear spin-lattice relaxation rate $1/T_1$ to explore the low-energy magnetic excitations in the ordered state. 
As shown in Fig.~\ref{relaxation}, $1/T_1$ becomes anisotropic below $T^{\ast}$. 
In $H_{\perp}$ $1/T_{1}$ shows a peak at $T^{\ast}$ due to the critical fluctuations near the magnetic phase transition and 
a power-law behavior was observed in the ordered state. 
The exponent close to $3$ is explained by the linear spin-wave theory for the $Q=0$ state,
whereas in $H_{||}$ the peak at $T^{\ast}$ is suppressed and the temperature dependence below $T^{\ast}$ is weaker than $T^3$. 
We suggest that the weak temperature dependence is due to the additional contribution in $H_{||}$, 
which corresponds to the $T$-linear contribution observed in the powder sample and also in the heat capacity measurement. \cite{ihara-PRB96, yoshida-JPSJ86}

The nuclear spins excited by the rf fields are relaxed by the local magnetic fluctuations perpendicular to the external field direction. 
The $T^3$ temperature dependence observed in $H_{\perp}$ strongly indicates that spin fluctuations in the kagome plane are 
perfectly understood as the dispersive magnon excitations, 
while the spin fluctuations along the $c$ direction have larger weight at lower energy, as sensed by $1/T_1$ in $H_{||}$. 
We address the zero mode as the origin of the low-energy excitations perpendicular to kagome plane. 
The finite and temperature independent $c$-axis susceptibility is also consistently explained by the zero mode. 
We should note that the flat band is lifted by the DM interaction, which results in a gapped excitations. 
The power-law temperature dependence was observed down to $2$ K in Ca-kapellasite 
because the gap by DM interaction is suppressed by $J_2$ and $J_d$ which introduce dispersion to the flat band. \cite{chernyshev-PRB92, matan-PRL96}

In Fig.~\ref{relaxation}, $1/T_1$ measured at $24$ T is shown together with the result at low field. 
In $H_{\perp}$ we observed a peak in $1/T_1$ at $T^{\ast}$ and $T^3$ temperature dependence below $T^{\ast}$ even at $24$ T. 
The peak temperature increases to $9$ K in $24$ T probably due to the weakly ferromagnetic nature of $Q=0$ state. 
To reveal the origin of the increase in $T^{\ast}$ in high magnetic fields, further experiments are necessary. 
In $H_{||}$ the peak at $T^{\ast}$ is strongly suppressed, and the $T^3$ behavior appears in a narrow temperature range just below $T^{\ast}$. 
Then the deviation from $T^3$ behavior was observed below $5$ K. 
From these results we suggest that the zero mode is more strongly suppressed by fields compared with the dispersive spin-wave contribution
because of the small energy scale relevant to the flat band. 
The heat capacity measurement also suggests that the $T$-linear contribution is suppressed by fields of about $10$ T. \cite{yoshida-JPSJ86}

\begin{figure}[tbp]
\begin{center}
\includegraphics[width=8.5cm]{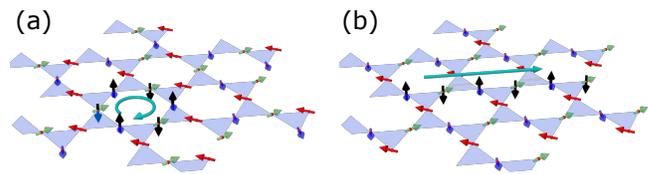}
\end{center}
\caption{
Real space images of zero modes in (a) $\sqrt{3} \times \sqrt{3}$ and (b) $Q=0$ structures. 
The spin excitations are localized around a hexagon in $\sqrt{3} \times \sqrt{3}$ structure, while they form a chain in $Q=0$ structure. 
} 
\label{Fig5}
\end{figure}

The zero mode from the $\sqrt{3} \times \sqrt{3}$ state is understood as the local excitations of six spins surrounding one hexagon. \cite{harris-PRB45} [Fig.~\ref{Fig5} (a)]
Such local excitation without any long range propagation will contribute to a temperature independent $1/T_1$ as in the case of free spins. 
In contrast, the zero mode from the $Q=0$ state forms a chain structure [Fig.~\ref{Fig5} (b)], 
which would introduce a power-law temperature dependence defined by the evolution of the coherence length. 
We propose that the low-energy excitations in Ca-kapellasite can be understood as the superposed one-dimensional chains weakly connected by long range interactions $J_2$, $J_d$. 
Further experiment is required to directly observe the low-energy magnon dispersion, and possibly additional continuum excitations originating from weakly dispersing low-energy band.

To summarize, we found the anisotropic spectrum broadening in the ordered state of Ca-kapellasite, 
which is consistently explained as the negative-chirality $Q=0$ state. 
From the $1/T_{1}$ measurement in $H_{||}$ and $H_{\perp}$ we conclude that
the in-plane component of the low-energy magnetic excitations are dominated by the spin-wave contribution. 
Moreover, we unveiled weaker temperature dependence for the magnetic excitations along the $c$ axis. 
These nontrivial excitations are consistent with the finite susceptibility along $c$ axis and also $T$-linear contribution to the heat capacity. 
We suggest that these excitations can be generated from the flat band in the kagome antiferromagnet when it obtains a weak dispersion due to long range interactions.

\begin{acknowledgements}
We would like to acknowledge Jun Ohara and Kazuki Iida for fruitful discussions. 
Part of this work was performed at the High Field Laboratory for Superconducting Materials, Institute for Materials Research, Tohoku University 
(Project Nos. 17H0046, 18H0015, 19H0037).
This study was partly supported by the Grant-in-Aid for Young Scientist B Grant Number JP15K17686 and JSPS KAKENHI Grant Numbers JP18K03529, JP19H01832, and  JP19H01846.
\end{acknowledgements}

\end{document}